\documentclass[journal abbreviation]{copernicus}

\begin{document}

\title{Acoustic oscillations in a field-free cavity under solar small-scale bipolar magnetic canopy}

\author[1]    {D. Kuridze}
\author[1]    {T. V. Zaqarashvili}
\author[1,2]  {B. M. Shergelashvili}
\author[3]    {S. Poedts}

\correspondence{David Kuridze// dato.k@genao.org}

\affil[1]{Georgian National Astrophysical Observatory (Abastumani
Astrophysical Observatory), Al. Kazbegi ave. 2a, 0160 Tbilisi,
Georgia} \affil[2]{Institute for Theoretical Physics, K.U. Leuven,
Celestijnenlaan 200 D, B-3001, Leuven, Belgium} \affil[3]{Center
for Plasma Astrophysics, K.U.Leuven, 200 B, B-3001, Leuven,
Belgium}
%% The [] brackets identify the author to the corresponding affiliation, 1, 2, 3, etc. should be inserted.

\runningtitle{Acoustic oscillations in a field-free cavity under
solar small-scale bipolar magnetic canopy}

\runningauthor{Kuridze et al.}

%\correspondence{dato.k@genao.org, temury@genao.org,
%Bidzina.Shergelashvili@wis.kuleuven.be,Stefaan.Poedts@wis.kuleuven.be}

\received{}
\pubdiscuss{} %% only important for two-stage journals
\revised{}
\accepted{}
\published{}

%% These dates will be inserted by the Publication Production Office during the typesetting process.

\firstpage{1}

\maketitle

\begin{abstract}

Observations show the increase of high-frequency wave power near
magnetic network cores and active regions in the solar lower
atmosphere. This phenomenon can be explained by the interaction of
acoustic waves with a magnetic field. We consider small-scale,
bipolar, magnetic field canopy structure near the network cores and
active regions overlying field-free cylindrical cavities of the
photosphere. Solving the plasma equations we get the analytical
dispersion relation of acoustic oscillations in the field-free
cavity area. We found that the $m=1$ mode, where $m$ is azimuthal
wave number, cannot be trapped under the canopy due to energy
leakage upwards. However, higher ($m \geq 2$) harmonics can be
easily trapped leading to the observed acoustic power halos under
the canopy.
\end{abstract}

\keywords{Sun: photosphere - Sun: acoustic oscillations}

\introduction
%% \introduction[modified heading if necessary]

Waves play an important role in the dynamics of the solar
atmosphere. Observations show an increase of high-frequency power
($\nu > 5\;$mHz) in the surroundings of active regions in velocity
power maps sometimes called as photospheric power halos
\citep{braun1992,brown1992,hindbro1998,jainheb2002}. The halos were
not found in Doppler power maps at lower frequencies ($3\;$ mHz).
Observations also show a lack of power halos in continuum intensity
power  maps \citep{hindbro1998,jainheb2002, Mug2005}.

On the other hand, recent observations reveal the decrease of the
acoustic high frequency power in the chromosphere and its increase
in the photosphere near active regions \citep{Mug2003,Mug2005}. It
has also been  shown that the quiet-Sun chromospheric magnetic
network elements are surrounded by "magnetic shadows", which lack
the oscillatory power at higher frequency range
\citep{maci2001,kru2001,Vecchio2007}. Therefore, both the
photospheric power halos and the chromospheric magnetic shadows
probably reflect the same physical process of acoustic wave
interaction with overlying magnetic field \citep[while in subsurface
regions the rotational and the meridional non-uniform flows are
supposed to have an impact on the formation of the acoustic wave
power spectra.][]{sherg2005}.

The properties of propagating acoustic waves are closely related to
the magnetic field structure. The numerical calculations show that
the propagation of acoustic disturbances in the solar atmosphere is
strongly determined by the overlying magnetic canopy
\citep{Rosenthal2002,Bogdan2003}.  The canopy has been usually
modeled with purely horizontal magnetic field \citep{Evans1990}, but
recent high-resolution observations reveal more complex small-scale
structure of the field \citep{dew,centeno2007}. It has been
suggested that the magnetic field has small-scale closed loop
structure in the vicinity of network cores
\citep{maci2001,schr2003}. The inclined magnetic field may channel
low-frequency photospheric oscillations in the chromosphere/corona
\citep{DePont2004}.

Here we use a model of small-scale bipolar magnetic canopy near a
chromospheric network core and/or an active region. We suggest that
granular cells may form field-free cylindrical cavities under the
magnetic canopy due to the transport of magnetic flux towards
boundaries. These cavities may trap high-frequency acoustic
oscillations, while the lower-frequency harmonics may propagate
upwards in form of magneto-acoustic waves.

Sect. 2 gives the analytical approach, obtained dispersion relation
and resulting oscillation spectrum. Sect. 3 includes discussion and
comparison of theoretical findings to observations. Sect. 4 briefly
summarizes the results.

\section{The model}

We use the ideal magnetohydrodynamic (MHD) equations which can be
written in the following form
\begin{equation}
{{{\partial \bf B}}\over {\partial t}}={\nabla}{\times}({\bf
v}{\times}{\bf B}),
\end{equation}
\begin{equation}
{\rho}{{{\partial \bf v}}\over {\partial t}} + {\rho}({\bf
v}{\cdot}{\nabla}) {\bf v} = - {\nabla}p + {{1}\over
{4{\pi}}}({\nabla}{\times}{\bf B}){\times}{\bf B},
\end{equation}
\begin{equation}
{{{\partial {\rho}}}\over {\partial t}} + {\nabla}{\cdot}(\rho{\bf
v})=0,
\end{equation}
\begin{equation}
{{{\partial p}}\over {\partial t}}+ ({\bf v}{\cdot}{\nabla})p +
{\gamma}p{\nabla}{\cdot}{\bf v}=0,
\end{equation}
where ${\bf v}$ denotes the fluid velocity, $\bf B$ the magnetic
field, $p$ the pressure, $\rho$ the mass density, and $\gamma$ the
ratio of specific heats. Gravity effect is omitted from the
consideration for the sake of simplicity. With this simple model we
intend to estimate the potential of the proposed mechanism. Gravity
will be taken into account in future developments of the present
model.

\begin{figure}[t]
\vspace*{2mm}
\begin{center}
\includegraphics[width=7 cm]{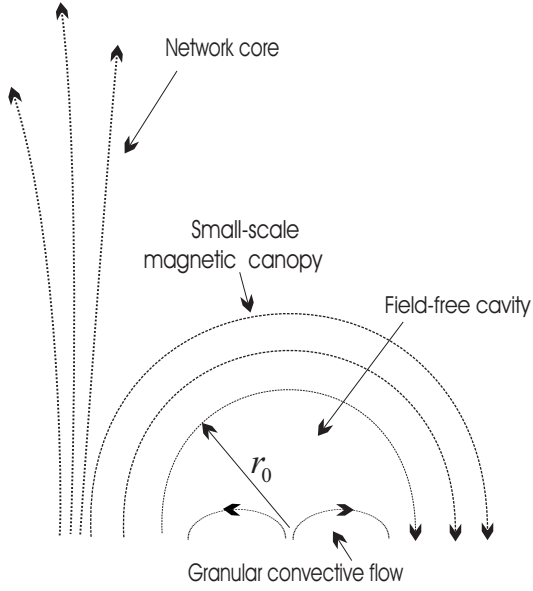}
\end{center}
\caption{Schematic picture of the magnetic canopy overlying a
field-free cavity.}
\end{figure}

We assume that the magnetic field is vertical in the  core of the
chromospheric network, but it gets the shape of small-scale closed
loop systems in the surrounding. \citep{maci2001,schr2003}. Granular
motions then may form field-free cavities under the canopy as they
carry the magnetic flux towards the cell boundaries. These
field-free cavities have granular dimensions and may become
resonators for acoustic oscillations. Consequently, we have two
different regions (Fig.~1):
\[ \left\{ \begin{array}{ll}
\mbox{region I}  - \mbox{field-free cavity,}  & r<r_0,  \\
\mbox{region II} - \mbox{magnetic canopy,} &r>r_0. \\
\end{array}
\right. \]

We use a cylindrical coordinate system $(r,\phi,z)$ and consider an
unperturbed cylindrical magnetic field $B_0$ in the canopy area. The
magnetic field has only a $\phi-$component which depends on the
distance \emph{r}, i.e.\ ${\bf B_0}=(0,B_{\phi}(r),0)$. The
equilibrium in the canopy is then satisfied if
$${d\over {dr}}\left ( p_0 + {{B_{\phi}^2}\over {8\pi}}\right ) +
{{B_{\phi}^2}\over {4{\pi}r}}=0,$$ where $p_0$ denotes the
unperturbed pressure. To avoid further mathematical complications,
we consider the unperturbed hydrodynamic pressure to be homogeneous.
The equilibrium magnetic field is then current-free expressed as
follows \citep{Diaz2006}
\begin{equation}
B_{\phi}=B_{\phi0}{r_0\over r}.
\end{equation}

Equations (1)-(4) are linearized, which yields
\begin{equation}
{{{\partial \bf b}}\over {\partial t}}={\nabla}{\times}({\bf
u}{\times}{\bf B_0}),
\end{equation}
$$
{\rho_0}{{{\partial \bf u}}\over {\partial t}} = -{\nabla}\left (p_1
+ {1\over {4\pi}}{{{\bf B_0}{\cdot}{\bf b}}}\right ) + {1\over
{4\pi}}({\bf B_0}{\nabla}){\bf b} +
$$
\begin{equation}
+{1\over {4\pi}}({{\bf b}{\nabla}}){\bf B_0},
\end{equation}
\begin{equation}
{{{\partial {\rho_1}}}\over {\partial t}} +
{\rho_0}{\nabla}{\cdot}{\bf u}=0,
\end{equation}
\begin{equation}
p_1={c_0^2}{\rho_1},
\end{equation}
where $\bf b$, $\bf u$, $p_1$ and $\rho_1$ are the perturbations of
magnetic field, velocity, pressure and mass density respectively,
while $\rho_0$ is a uniform unperturbed mass density and
$c_0=({\gamma}p_0/\rho_0)^{1/2}$ corresponds to the homogeneous
sound speed. Note, that a background flow is absent in our
consideration.

For simplicity, we consider the 2D case and restrict the analysis to
the $(r,\phi)-$plane. In principle, the $z-$direction can also be
considered, but this further complicates the presentation (namely,
resonant absorption may take place) and it is left for future
considerations. Then $r-$ and $\phi-$components of equations (6)-(9)
are given by
\begin{equation}
{{{\partial b_r}}\over {\partial t}}={{B_{\phi}}\over r}{{{\partial
u_r}}\over {\partial {\phi}}},
\end{equation}
\begin{equation}
{{{\partial b_{\phi}}}\over {\partial t}}= - B_{\phi}{{{\partial
u_r}}\over {\partial r}} + B_{\phi}{{u_r}\over r},
\end{equation}
\begin{equation}
\rho_0{{{\partial u_r}}\over {\partial t}}= -c_0^2{{\partial
{\rho_1}}\over {\partial r}} - {{B_{\phi}}\over {4\pi}}{{{\partial
b_{\phi}}}\over {\partial r}} + {{B_{\phi}}\over
{4{\pi}r}}{{{\partial b_r}}\over {\partial {\phi}}} -
{{B_{\phi}}\over {{4\pi}r}}b_{\phi},
\end{equation}
\begin{equation}
\rho_0{{{\partial u_{\phi}}}\over {\partial t}}= -{{c_0^2}\over
r}{{\partial {\rho_1}}\over {\partial {\phi}}},
\end{equation}
\begin{equation}
{{{\partial {\rho_1}}}\over {\partial t}} + \rho_0{{{\partial
u_r}}\over {\partial r}} + \rho_0{{u_r}\over r} + {{\rho_0}\over
r}{{\partial u_{\phi}}\over {\partial {\phi}}}=0.
\end{equation}

To get the oscillation spectrum in the cavity area, we have to solve
the equations in the cavity and canopy regions separately and then
merge the obtained solutions at the interface ($r=r_0$).

Medium is field-free in the cavity ($r<r_0$) and therefore can be
described by pure hydrodynamics. Then equations (12)-(14) lead to
the Bessel equation after Fourier analysis with respect to both the
\emph{t} (time) and $\phi$ coordinates (the magnetic field is set to
zero):
\begin{equation}
{{\partial^2 {\rho_1}}\over {\partial r^2}} + {1\over r}{{\partial
{\rho_1}}\over {\partial r}} + \left [{{\omega^2}\over c_0^2} -
{m^2\over {r^2}} \right ]{\rho_1}=0,
\end{equation}
where $\omega$ is the wave frequency, and $m$ is the azimuthal wave
number.

In the magnetized canopy region ($r>r_0$), equations (10)-(14) lead
to the Hain-Lust equation
$$
{{\partial}\over {\partial r}}\left ({{\omega^2c_0^2}\over
{\omega^2-m^2c_0^2/r^2}}+ v_A^2\right ){{\partial {\hat u_r}}\over
{\partial r}} + $$$$ +{{\partial}\over {\partial r}}\left
({{\omega^2c_0^2}\over {\omega^2-m^2c_0^2/r^2}} + v_A^2\right
){{{\hat u_r}}\over r} +
$$
\begin{equation}
+ \left [\omega^2 + {{4v_A^2}\over {r^2}} - {{m^2v_A^2}\over {r^2}}
\right ]{\hat u_r}=0,
\end{equation}
where $v_A=B_{\phi}/\sqrt{{4\pi}{\rho_0}}$ denotes the Alfv{\'e}n
speed.

Analytical solution of equation (16) is complicated. Therefore, for
simplicity, we suppose that the magnetic energy is much higher than
the hydrodynamic one within the canopy area. Consequently, we use
the zero $\beta$ approximation hereinafter. The unperturbed
configuration must be in equilibrium, therefore the unperturbed
hydrodynamic pressure of the cavity and the magnetic pressure of the
canopy must be balanced at the interface ($r=r_0$)
\begin{equation}
{c^2_{01} \rho_{01}\over \gamma}={B^2_{\varphi}(r_0)\over 8\pi},
\end{equation}
where
\begin{equation}
c_{01}=\sqrt{{\gamma p_0}\over \rho_{01}},
\end{equation}
and $c_{01}$ and $\rho_{01}$ are the sound speed and the plasma mass
density in the cavity. Equation (17) then gives the relation between
the sound and Alfv{\'e}n speeds at the interface
\begin{equation}
c_{01} = \sqrt{{\gamma \over 2}{\rho_{02}\over \rho_{01}}} v_{A2},
\end {equation}
where
\begin{equation}
v_{A2}= {B_{\varphi}(r_0)\over \sqrt{4\pi \rho_{02}}}
\end {equation}
is the Alfv{\'e}n speed at the interface and $\rho_{02}$ is the
plasma mass density in the canopy.

\subsection{Analytical Solutions}

In the field-free cavity area under the canopy (region~I), there are
only acoustic waves. The $r$-dependence of the mass density
perturbations in the acoustic waves is governed by equation (15)
which has the general solution
\begin{equation}
{\rho}_1=c_{1}J_m(k_1r) + c_{2}Y_m(k_1r),
\end{equation}
where $J_m(k_1r)$ ($Y_m(k_1r)$) is the bessel function of the first
(second) kind,
\begin{equation}
k_1={{\omega}\over {c_{01}}}
\end{equation}
and $c_{1}$, $c_{2}$ are arbitrary constants. The perturbation must
be finite at $r=0$. Therefore, $c_{2}=0$ and only the first term on
the right hand side of expression (21) is non-vanishing.

The radial velocity component of the acoustic wave in the region~I
can be obtained using equations (12) and (21) as (setting the
magnetic field to be zero)
\begin{equation}
{\hat u_{r1}}={{ic^2_{01}}\over {\rho_{01}
\omega}}c_{1}J^{\prime}_m(k_1r),
\end{equation}
where the prime denotes the derivation with respect to  $r$.

In the magnetic canopy (region~II), where the cold plasma
approximation is used, equation (16) gives
\begin{equation}
{{\partial^2 {\hat u_r}}\over {\partial r^2}}-{1\over r}{{\partial
{\hat u_r}}\over {\partial r}} + \left [{\omega^2\over v_A^2(r)} +
{1\over {r^2}} - {m^2\over {r^2}} \right ]{\hat u_r}=0.
\end{equation}
This equation can be rewritten as
\begin{equation}
{1\over r}{{\partial}\over {\partial r}}\left (r{{\partial
{\psi}}\over {\partial r}}\right ) + \left [{{\omega^2}\over
v_A^2(r)} - {m^2\over {r^2}} \right ]{\psi}=0,
\end{equation}
where $${\hat u_r}=r\psi,\,\, v_A^2(r)={{B^2_{\phi0}}\over {4\pi
\rho_{02}}}{r^2_0\over r^2}.$$

The solutions of this equation are the Bessel functions of half
integer order \citep{AbramSteg1967, Diaz2006}. In region~II, only
the outgoing wave is physically appropriate. Therefore, we choose
the Hankel function
\begin{equation}
{\hat u_{r2}}=ic_3rH_{m/2}(k_2r),
\end{equation}
where $c_3$ is an arbitrary constant and
\begin{equation}
k_2={{\omega}\over {2v_{A2}}}.
\end{equation}

The total pressure perturbations in the regions I and II are, respectively,
\begin{equation}
p_{1}=c^2_{01}{\rho_1},
\end{equation}
and
\begin{equation} {{B_{\phi}b_{\phi}}\over {4\pi}}= {{i\rho_{02}v^2_{A2}}\over {\omega}}\left [{{d{\hat
u_{r2}}}\over dr} - {{\hat u_{r2}}\over r}\right ].
\end{equation}

Thus the expressions (23), (26), (28), and (29) give the
transverse velocity and the total pressure perturbations in the
considered two regions.

\subsection{Dispersion relation}

The continuity of the velocity and the total pressure
perturbations at the interface ($r=r_{0}$) leads to
\begin{equation}
c^2_{01} {\rho}_1 = {{i\rho_{02}v^2_{A2}}\over {\omega}}\left
[{{d{\hat u_{r2}}}\over dr} - {{\hat u_{r2}}\over r}\right ],
\end{equation}
\begin{equation}
\hat u_{r1}=\hat u_{r2}.
\end{equation}

The substitution of the expressions ${\rho}_1, \hat u_{r1}, \hat
u_{r2}$ into equations (30) and (31) then gives
\begin{equation}
c^2_{01} c_1 J_m(k_1r_0)= -{{\rho_{02}v^2_{A2}}\over {\omega}}c_3
r_o k_2 H_{m/2}^{\prime}(k_2r_0),
\end{equation}
\begin{equation}
{{c^2_{01} }\over {\omega \rho_{01}}}c_1 k_1
J_m^{\prime}(k_1r_0)=c_3 r_o H_{m/2}(k_2r_0).
\end{equation}

The condition for a non-trivial solution of equations (32)-(33),
then yields the general dispersion relation, viz.\
\begin{equation}
{{\omega^2}\over v^2_{A2}}{\rho_{01}\over
\rho_{02}}{{J_m(k_1r_0)}\over k_1 J_m^{\prime}(k_1r_0)}= -{{k_2
H_{m/2}^{\prime}(k_2r_0)}\over H_{m/2}(k_2r_0)}.
\end{equation}

\begin{figure}[t]
\vspace*{2mm}
\begin{center}
\includegraphics[width=9.4 cm]{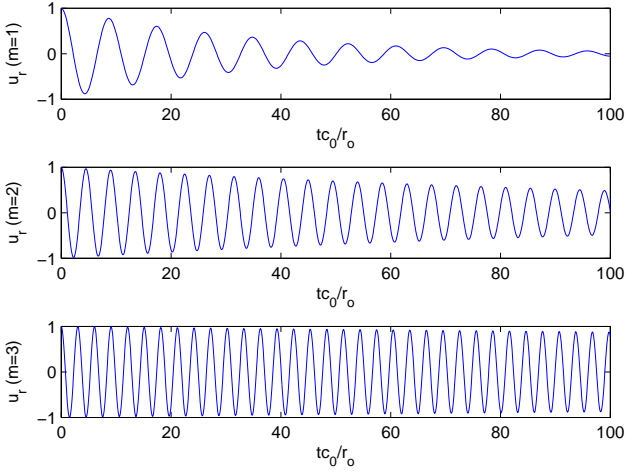}
\end{center}
\caption{Temporal dynamics of velocity perturbations in the
field-free cavity areas for $m=1,2,3$ harmonics.}
\end{figure}

The dispersion relation (34) is a transcendental equation for the
complex $\omega$. The imaginary part of wave frequency $\omega$
indicates a wave leakage from the field-free cavity into the ambient
magnetic canopy. The question then arises how important this wave
leakage is and which are the most sensitive parameters for trapping
(or leakage) of the waves under the canopy. The analytical solution
of equation (34) is complicated. Therefore, we apply numerical
techniques to solve it.

The properties of the dispersion relation (34) depend on the
azimuthal wave number $m$ and the ratio between sound ($c_{01}$) and
Alfv{\'e}n ($v_{A2}$) speeds (or the ratio between $\rho_{01}$ and
$\rho_{02}$, see Eq.~(19)). The dimension of the field-free cavity,
$r_0$, also stands as a free parameter, but it influences only the
wave periods.

Numerical solution of equation (34) shows that the ratio of
imaginary $\omega_i$ and real $\omega_r$ parts significantly depends
on the azimuthal wave number $m$. The ratio is higher for $m=1$ and
quickly decreases with increasing $m$. Fig.~2 shows the temporal
dynamics of radial velocity perturbations in the field-free cavity
region for modes corresponding to different $m$-values. The
amplitude of the first ($m=1$) harmonic quickly decreases, which
indicates the rapid radiation of this mode into the overlying
canopy. On the other hand, the $m=2$ mode undergoes a very small
leakage and the $m=3$ mode has almost no leakage. Thus, the first
harmonic of the acoustic oscillations cannot be trapped in the
cavity area due to the rapid leakage into the canopy, while the
higher harmonics, i.e.\ those with $m \geq 2$, can be easily
trapped, which may lead to the observed increased acoustic power.

Fig.~3 shows the ratio of the imaginary $\omega_i$ and real
$\omega_r$ parts vs the ratio of the Alfv{\'e}n and sound speeds for
different harmonics. We see that the decrease of the ratio between
the Alfv{\'e}n and sound speeds leads to an enhanced leakage.

\begin{figure}[t]
\vspace*{2mm}
\begin{center}
\includegraphics[width=9.4 cm]{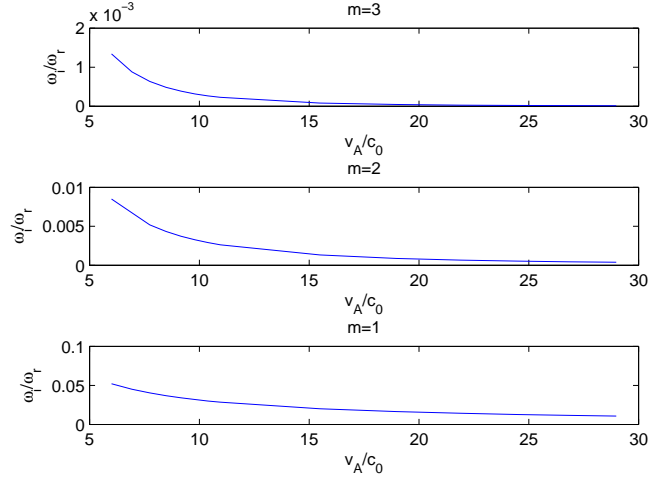}
\end{center}
\caption{The ratio of imaginary $\omega_i$ and real $\omega_r$ parts
of wave frequency vs the ratio of Alfv{\'e}n and sound speeds for
$m=1,2,3$ harmonics.}
\end{figure}

Fig.~4 shows the dependence of real (top panel) and imaginary
(bottom panel) parts of the frequency on the size $r_0$ of the
field-free cavity region under the canopy. It is evident that the
real part of the frequency decreases with increasing $r_0$, as can
be expected from physical considerations.

\section{Discussion}

We have studied the spectrum of acoustic oscillations in a
cylindrical field-free cavity under a small-scale bipolar magnetic
canopy in the solar atmosphere. It is shown that the $m=1$ (where
$m$ is the azimuthal wave number) harmonic of the acoustic
oscillations cannot be trapped in the cavity as a result of the
energy leakage in the upward direction. The energy radiation occurs
through the propagation of fast magneto-acoustic waves in overlying
magnetic canopy.
%\citep{Verwichte2006, Diaz2006}%
However, higher $m \geq 2$ harmonics can be trapped in the cavity,
leading to the observed increased high-frequency power in the
photosphere.

%As a matter of fact, observations show an increase of high-frequency
%power ($\nu > 5\;$mHz) in the surroundings of active regions
%(photospheric power halos) in MDI velocity power maps
%\citep{hindbro1998,jainheb2002, Mug2005}. The halos were not found
%in Doppler power maps at lower frequencies ($3\;$mHz). Observations
%also show a lack of power halos in continuum intensity power  maps
%\citep{hindbro1998,jainheb2002, Mug2005}.%

There are three different explanations of power halos proposed in
the literature: (i)~the enhancement of acoustic emission by some
unknown source \citep{braun1992,brown1992,jainheb2002},
(ii)~incompressible oscillations, such as Alfv{\'e}n waves or
transverse kink waves, in magnetic tubes \citep{hindbro1998}, and
(iii)~the interaction of acoustic waves with the overlying magnetic
canopy \citep{Mug2005}.

We suggest that the surroundings of magnetic network cores and
active regions consist of many small-scale closed magnetic canopy
structures \citep{maci2001,schr2003}. Granular motions transport the
magnetic field at boundaries and consequently create field-free
cylindrical cavity areas under the canopy (see Fig.~1). These
field-free cavities may be filled by trapped acoustic oscillations
stochastically excited by granular motions \citep{Lighthill1952,
Selwa2004}. For a typical photospheric sound speed of $\sim7\;$km/s
and a typical granular radius of $\sim400\;$km, the period of the
$m=2$ harmonic is $\sim3\;$min and the period of the $m=3$ harmonic
is $\sim2\;$min. The range of observed enhanced power frequency is
about $5-7\;$mHz (with periods of $2-3\;$min), which is in agreement
with our findings. The $m=1$ harmonic is leaky and, therefore, can
not be trapped under the canopy.

If power halos are related to the acoustic waves then observations
should show an enhancement in continuum intensity power maps as
well, but in opposite observations show a lack of power halos in the
maps \citep{hindbro1998, jainheb2002, Mug2005}, what needs an
adequate explanation.
%%% I do not understand the above sentence- corrected
Here, we suggest a natural explanation for this phenomenon. Indeed,
the antinodes of mass density and velocity components in standing
acoustic oscillations are located at different places: the location
of the maximal velocity oscillations corresponds to the location of
the zero amplitude intensity oscillations and vice versa. Therefore,
the enhanced Doppler velocity power at fixed height of the
atmosphere automatically suggest the absence of intensity power at
the same height. This suggestion can be checked by searching
intensity power halos at different heights from the surface. We
believe that new observations with high resolution from the
\emph{Hinode} spacecraft will shed light on this problem.

\begin{figure}[]
\vspace*{2mm}
\begin{center}
\includegraphics[width=9.4 cm]{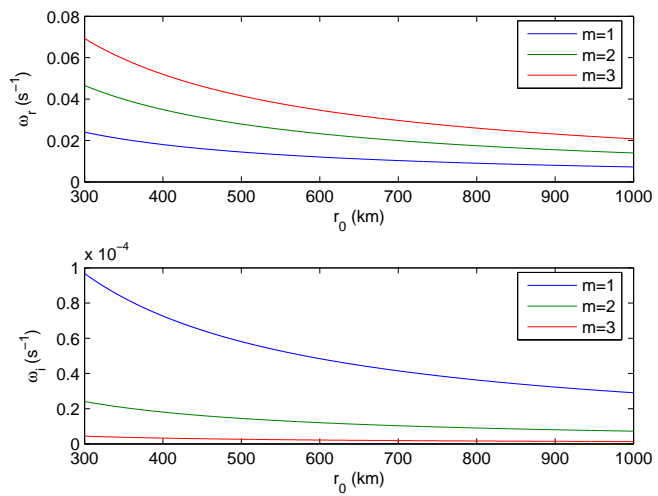}
\end{center}
\caption{Real $\omega_r$ (upper panel) and imaginary $\omega_i$
(lower panel) parts of wave frequency vs the radius of field-free
cavity $r_0$ for $m=1,2,3$ harmonics}.
\end{figure}

However, the formation of power halos due to incompressible
oscillations can not be completely ruled out. Recent theoretical and
two-dimensional numerical simulations outlined the importance of the
$\beta \sim 1$ region in the solar atmosphere, where wave conversion
or reflection occurs \citep{Rosenthal2002,Bogdan2003,ZaqRob2006,
KuriZaq2007}. Indeed, our consideration implies that the
hydrodynamic and magnetic pressures have approximately the same
value at the interface (see Eq.~17). Therefore, the acoustic waves
may transfer energy into the incompressible waves through non-linear
interactions \citep{ZaqRob2006,KuriZaq2007}, which may lead to the
observed power halos. However, the observed oscillations hardly show
non-linear behavior, which complicates the explanation of power
halos by incompressible oscillations considerably.

Another important alternative for the proposed mechanism can be
resonant absorption of waves in inhomogeneous plasma
\citep{tirry1998,pintgoos1999,pinter2007}. This process can also be
particularly important for 3D consideration of proposed model, when
one considers the wave propagation in the $z$ direction as well.
This is out of the scope of present paper, but would be interesting
to study in the future.

It must be mentioned that the equilibrium used in this paper is
simplified as gravitational stratification, which is important in
the solar atmosphere, is ignored. The stratification leads to a
Klein-Gordon equation for the propagating waves with a cut-off for
wave frequencies \citep{Roberts2004,Erdelyi2007}. However, wave
propagation in inclined magnetic field may lead to decrease of
cut-off frequencies and simplifies the penetration of
lower-frequency oscillations in higher regions \citep{DePont2004}.
Non of the harmonics has purely vertical propagation in our model
because of the non-zero azimuthal wave number $m$; all harmonics
propagate with an angle to the vertical (note, that all harmonics
have standing wave behavior along azimuthal direction). Therefore
their frequencies are above the cut-off value and consequently the
harmonics are not evanescent. However, inclusion of gravitational
stratification is necessary for a more profound understanding of the
wave trapping in cavities. This will be the next step of our study.

%We plan to study the similar problem with the gravitational
%stratification for a more profound understanding of the complete
%picture of wave trapping.

\section{Conclusion}

We have studied the spectrum of acoustic oscillations in the
cylindrical field-free cavity regions under the small-scale magnetic
canopy near the magnetic network cores and active regions. We found
that the first harmonic of acoustic oscillations cannot be trapped
in the cavity due to the energy radiation by fast magneto-acoustic
waves in the canopy. However, the higher ($m \geq 2$) harmonics can
be trapped there, leading to the observed enhancement of
high-frequency acoustic power in the photosphere. Future detailed
study of the proposed mechanism including gravitational
stratification and 3D models is necessary.

\begin{acknowledgements}
The work was supported by the grant of Georgian National Science
Foundation GNSF/ST06/4-098. The work of B.M.S. has been supported
by K.U.Leuven scholarship - PDM/06/116. These results were
obtained in the framework of the projects GOA/2004/01
(K.U.Leuven), G.0304.07 (FWO-Vlaanderen) and C~90203 (ESA Prodex
8). Financial support by the European Commission through the
SOLAIRE Network (MTRN-CT-2006-035484) is gratefully acknowledged.
\end{acknowledgements}


\begin{thebibliography}{}



\bibitem[Abramowitz \& Stegun(1967)]{AbramSteg1967}
Abramowitz, M., and Stegun I. A.: Handbook of Mathrmatical Functions
(Dover), 1967

\bibitem[Bogdan et al.(2003)]{Bogdan2003}
Bogdan, T. J., Hansteen, M., Carlsson, V., et al.: Waves in the
magnetized solar atmosphere. II. Waves from localized sources in
magnetic flux concentrations, ApJ, 599, 626-660, 2003

\bibitem[Braun et al.(1992)]{braun1992}
Braun, D. C., Lindsey, C., Fan, Y., and
Jefferies, S. M.: Local acoustic diagnostics of the solar
interior, ApJ, 392, 739-745, 1992

\bibitem[Brown et al.(1992)]{brown1992}
Brown, T. M, Bogdan, T. J., Lites, B. W., and
Thomas, J. H.: Localized sources of propagating acoustic waves in
the solar photosphere, ApJ, 394, 65-68, 1992

\bibitem[Centeno et al.(2007)]{centeno2007}Centeno, R.,
Socas-Navarro, H., Lites, B., Kubo, M., Frank, Z., Shine, R.,
Tarbell, T., Title, A., Ichimoto, K., Tsuneta, S., Katsukawa, Y.,
Suematsu, Y., Shimizu, T., \& Nagata, S.:Emergence of Small-Scale
Magnetic Loops in the Quiet-Sun Internetwork, ApJ 666, 137-140, 2007

\bibitem[De Pontieu et al.(2004)]{DePont2004}
De Pontieu, B., Erdelyi, R., and E., James, S. P.: Solar
chromospheric spicules from the leakage of photospheric oscillations
and flows, Nature, 430, 536-539, 2004

\bibitem[De Wijn et al.(2005)]{dew}De Wijn, A.G., Rutten, R.J., Haverkamp, E.M.W.P., and Sutterlin, P.,
2005, A\&A, 441, 1183

\bibitem[D\'iaz et al.(2006)]{Diaz2006}
D\'iaz, A. J., Zaqarashvili, T., and Roberts, B.: Fast
magnetohydrodynamic oscillations in a force-free line-tied coronal
arcade, A\&A, 455, 709-717, 2006

\bibitem[Erd\'elyi et al.(2007)]{Erdelyi2007} Erdelyi, R., Malins,
C., Toth, G., and de Pontieu, B.: Leakage of photospheric acoustic
waves into non-magnetic solar atmosphere, A\&A, 477, 1299-1311, 2007

\bibitem[Evans \& Roberts(1990)]{Evans1990}Evans, D.J., and
Roberts, B.: The influence of a chromospheric magnetic field on the
solar p- and f-modes. II - Uniform chromospheric field, ApJ, 356,
704-719, 1990

\bibitem[Hindman \& Brown(1998)]{hindbro1998}
Hindman, B. W., and Brown, T. M.: Acoustic power maps
of solar active regions, ApJ, 504, 1029, 1998


\bibitem[Jain \& Haber(2002)]{jainheb2002}
Jain, R., and Haber, D.: Solar p-modes and surface
magnetic fields: Is there an acoustic emission?. MDI/SOHO
observations, A\&A, 387, 1092-1099, 2002

\bibitem[Krijger et al.(2001)]{kru2001}Krijger, J.M., Rutten, R.J., Lites, B.W., Straus, Th., Shine, R.A. \& Tarbell,
  T.D.: Dynamics of the solar chromosphere. III. Ultraviolet brightness oscillations from TRACE, A\&A, 379,
  1052, 2001

\bibitem[Kuridze \& Zaqarashvili(2007)]{KuriZaq2007}
Kuridze, D. and Zaqarashvili, T. V. :Resonant energy conversion of
3-minute intensity oscillations into Alfven waves in the solar
atmosphere, JASTP, (accepted), 2007


\bibitem[Lighthill(1952)]{Lighthill1952}
Lighthill, M. J.: On sound generated aerodynamically. I. General
theory, Proc. Roy. Sos., 211, 564-587, 1952

\bibitem[McIntosh \& Judge(2001)]{maci2001}McIntosh, S. W. \&
Judge, P.G.,: On the Nature of Magnetic Shadows in the Solar
Chromosphere, ApJ, 561, 420, 2001

\bibitem[Muglach(2003)]{Mug2003}
Muglach, K.: Dynamics of solar active regions. I. Photospheric and
chromospheric oscillations observed with TRACE,  A\&A, 401, 685-697,
2003

\bibitem[Muglach et al.(2005)]{Mug2005}
Muglach, K., Hofmann, A., and Staude, J.: Dynamics of solar active
regions. II. Oscillations observed with MDI and their relation to
the magnetic field topology,  A\&A, 437, 1055-1060, 2005

\bibitem[Pint\'er et al.(2007)]{pinter2007} Pint\'er, B.; Erdelyi,
R., and Goossens, M.:Global oscillations in a magnetic solar model.
II. Oblique propagation, A\&A, 347, 321-334 , 1999

\bibitem[Pint\'er \& Goossens(1999)]{pintgoos1999} Pint\'er, B.,
and  Goossens, M.: Oscillations in a magnetic solar model. I.
Parallel propagation in a chromospheric and coronal magnetic field
with constant Alfven speed, A\&A, 466, 377-388, 2007

\bibitem[Roberts(2004)]{Roberts2004}
Roberts, B.: MHD Waves in the Solar Atmosphere, In Proc. of 'SOHO 13
- Waves, Oscillations and Small-Scale Transient Events in the Solar
Atmosphere: A Joint View from SOHO and TRACE', Palma de Mallorca,
Balearic Islands, Spain (ESA SP-547), 1, 2004

\bibitem[Rosenthal et al.(2002)]{Rosenthal2002}
Rosenthal, C. S., Bogdan, T. J., Carlsson, M., Dorch, S. B. F.,
Hansteen, V., McIntosh, S. W., McMurry, A., Nordlund, and Stein, R.
F.: Waves in the magnetized solar atmosphere. I. Basic processes and
internetwork oscillations, ApJ 564, 508-524, 2002

\bibitem[Selwa et al.(2004)]{Selwa2004}
Selwa, M., Skartlien, R., and Murawski, K.: Numerical simulations of
stochastically excited sound waves in a random medium, A\&A, 420,
1123-1127, 2004

\bibitem[Shergelashvili \& Poedts(2005)]{sherg2005}
Shergelashvili, B. M. and Poedts, S.: On the effect of the
inhomogeneous subsurface flows on the high degree solar p-modes,
A\&A, 438, 1083-1097, 2005

\bibitem[Schrijver \& Title(2003)]{schr2003}Schrijver, C.J. \&
Title, A.M.:The Magnetic Connection between the Solar Photosphere
and the Corona, ApJ, 597, L165, 2003

\bibitem[Tirry et al.(1998)]{tirry1998} Tirry, W. J., Goossens,
M., Pinter, B., Cadez, V., and Vanlommel, P.: Resonant Damping of
Solar p-Modes by the Chromospheric Magnetic Field, ApJ, 503, 422,
1998

\bibitem[Vecchio et al.(2007)]{Vecchio2007}
Vecchio, A., Cauzzi, G., Reardon, K. P., Janssen, K., and Rimmele,
T.: Solar atmospheric oscillations and the chromospheric magnetic
topology, A\&A, 461, 1-4, 2007

\bibitem[Zaqarashvili \& Roberts(2006)]{ZaqRob2006}
Zaqarashvili, T.V. and Roberts, B.: Two-wave interaction in ideal
magnetohydrodynamics, A\&A, 452, 1053-1058, 2006
\end{thebibliography}
\end{document}